\documentclass[prl,aps,twocolumn,footinbib,showpacs]{revtex4}
\usepackage{bm,graphicx,amsmath,bbm}
\usepackage{txfonts}

\newcommand{\abs}[1]{\left|{#1}\right|}

\newcommand{\av}[1]{\left\langle #1 \right\rangle}

\newcommand{\br}[1]{\langle #1|}

\newcommand{\ke}[1]{|#1\rangle}

\newcommand{\kb}[2]{\ke{#1}\br{#2}}

\newcommand{\al}[1]{^{(#1)}}
\newcommand{\da}{^\dagger}

\newcommand{\pt}[1]{\left( #1 \right)}
\newcommand{\pq}[1]{\left[ #1 \right]}
\newcommand{\pg}[1]{\left\{ #1 \right\}}

\newcommand{\bs}[1]{\boldsymbol #1}

\newcommand{\ee}{{\rm e}}
\newcommand{\ii}{{\rm i}}
\newcommand{\dd}{{\rm d}}

\newcommand{\id}{\mathbbm{1}}

\newcommand{\nn}{{\nonumber}}

\begin{document}

\title{Entanglement replication in driven-dissipative many body systems}
\author{S. Zippilli$^1$, M. Paternostro$^{2,3}$, G. Adesso$^4$, and F. Illuminati$^{1,}$\footnote{Corresponding author: illuminati@sa.infn.it}}
\affiliation{
$^1$\mbox{Dipartimento di Ingegneria Industriale, Universit\`a degli Studi di Salerno,~Via Ponte don Melillo, I-84084 Fisciano (SA), Italy}
\\
$^2$\mbox{Centre for Theoretical Atomic, Molecular, and Optical Physics, School of Mathematics and Physics, Queen's University, Belfast BT7 1NN, UK}\\
$^3$\mbox{Institut f\"ur Theoretische Physik, Albert-Einstein-Allee 11, Universit\"at Ulm, D-89069 Ulm, Germany}\\
$^{4}$\mbox{School of Mathematical Sciences, University of Nottingham,\,University\,Park, Nottingham NG7 2RD, United Kingdom}
}

\date{December 6, 2012}

\begin{abstract}
We study the dissipative dynamics of two independent arrays of many-body systems, locally driven by a common entangled
field. We show that in the steady state the entanglement of the driving field is reproduced in an arbitrarily large series of inter-array entangled pairs over all distances. Local nonclassical driving thus realizes
a scale-free
{\it entanglement replication} and long-distance entanglement distribution mechanism that has immediate bearing on the implementation of quantum communication networks.
\end{abstract}
\pacs{03.67.Bg, 42.50.Dv, 03.65.Yz, 42.50.-p}

\maketitle

Driving quantum systems to desired target states with very high fidelity is a central goal in quantum sciences and technologies, in order to realize efficient
and scalable devices beyond the current state of proof-of-principle demonstrations. In pursuing this end,
it has surfaced in recent years that the effects of noise and dissipation do not have to be necessarily detrimental in the realization of quantum coherent structures~\cite{Poyatos,Plenio2002,Benatti2003,Pielawa,Schmidt2011}.
The possibility of using suitably engineered
irreversible dynamics to  control quantum many-body systems
has been discussed in a variety of settings, including driven-dissipative
ultracold atoms in optical lattices \cite{Diehl2008},
the asymptotic realization of entangled states and quantum computation
in quantum spin models~\cite{Kraus2008,Verstraete2009},
the dissipative control of trapped ions~\cite{Barreiro},
and the steady-state entanglement of macroscopic atomic ensembles~\cite{Krauter2011}.
On the other hand, ever since the formulation of the proposal for quantum repeaters~\cite{repeaters} and the design of schemes for the implementation of remote quantum communication and distributed quantum gates~\cite{nonlocal}, quantum networks have emerged as the strongest viable paradigm for the "quantum internet", i.e. the implementation of scalable quantum computation and information processing satisfying the combined requirements of robustness, flexibility,
multi-tasking and long-reach~\cite{kimble}.
A key ingredient of a quantum internet is the ability to hybridize, i.e. to interface heterogeneous subsystems in a reliable and reproducible way. The strive towards the realization of such interfaces has been boosted by recent ground-breaking demonstrations of high-efficiency entanglement and state transfer between light and matter systems~\cite{light-matter,blinov,Togan},
and light-mediated teleportation between remote nodes of a simple quantum network~\cite{monroe}.
\begin{figure}[b]
\centering
\includegraphics[width=8cm]{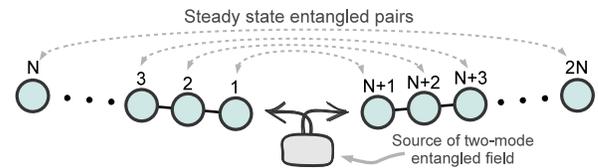}%
\caption{
A pair of independent arrays of linearly coupled quantum systems is locally driven by a two-mode entangled field. The elements in each array are labeled by the indices $j{\in}\pq{1,N}$ (first chain) and $j{\in}\pq{N{+}1,2N}$ (second chain). The steady-state inter-array entangled pairs are marked by dashed arrows.
}
\label{fig1}
\end{figure}

In this context, light-matter interfaces for the distribution of entanglement among network nodes, which exploit the robustness of irreversible dynamics, have been explored in several works~\cite{kraus,paternostro2004,adesso10}. There it was shown	that a reservoir of entangled light can drive distant matter systems into entangled states, thereby realizing an efficient transfer of entanglement from continuous- to discrete-variable systems.

In the present work we show that when considering independent arrays of many-body quantum systems this mechanism amounts to the {\it replication} of the driving entanglement over many pairs of subsystems across the initially independent arrays. Specifically, we address the irreversible dynamics of two non-interacting chains of quantum systems simultaneously driven, on one of their ends, by an entangled two-mode squeezed field (squeezed bath). The constituents in each array are coupled by nearest-neighbor linear interactions whose specific form is introduced below for different models.
The competition between the ``entanglement pumping'' process and the intra-array couplings results in a steady state consisting of a series of inter-array entangled pairs, each involving subsystems occupying corresponding sites in the respective chain [See Fig.~\ref{fig1}].
Thereby, an arbitrary number of copies of identically entangled states is generated across the two arrays  without violating fundamental constraints such as the no-cloning and the no-broadcasting theorems~\cite{nocloning}.

The replication mechanism works efficiently in different settings such as chains of harmonic oscillators or of spins. For pure harmonic resonators in the stationary state exactly $N$ inter-chain pairs are formed that replicate the driving state independently of the size of the arrays. For two-level systems an ideal Einstein-Podolsky-Rosen (EPR) driving field creates exactly $N$ Bell states across the two chains.

To start, let us consider two chains of resonators, realizing two disjoint Jaynes-Cummings lattices~\cite{JCH1,JCH2}, which can describe, in limiting cases, the physics of different condensed-matter systems ranging from spin chains, to boson or fermion lattice models.
The two arrays are assumed equal (deviations from this condition are discussed below), and each consists of $N$ single-mode cavities with equal resonance frequency and corresponding annihilation (creation) operators $\hat a_j$ ($\hat a_j\da$). Cavities belonging to the same array interact via nearest-neighbor linear coupling with strength $\eta_j$. Moreover, each cavity can interact resonantly with a two-level system (e.g.~an atom in the cavity) with lowering (rising) operator  $\hat\sigma_j$ ($\hat\sigma_j\da$).
As illustrated in Fig.~\ref{fig1}, the elements of the first (second) array are labeled by indices $j\in[1,N]$ ($j\in[N+1,2N]$). The two end cavities $1$ and $N+1$ are driven by a two-mode squeezed field. Including the dissipation of the cavity fields \cite{se}, the master equation describing the system dynamics is
$\dot\rho=-i [{ H_c+ H_{cs},\rho}]+{\cal L}_D\rho+{\cal L}_S\rho .$
The unitary part of the evolution is ruled by the Hamiltonian $ H_c{+} H_{cs}$ with $ H_{c}{=}\sum_{j=1}^{N-1}\eta_j(\hat a_j\da \hat a_{j+1}+\hat a_{N+j}\da \hat a_{N+j+1}+\text{h.c.})$ describing the coherent cavity dynamics and $ H_{cs}{=}\sum_{j=1}^{N}g_j(\hat \sigma_j\da \hat a_j+\hat \sigma_{N+j}\da \hat a_{N+j}+\text{h.c.})$ accounting for the interaction (with coupling $g_j$) between cavity $j$ and its two-level system. The term ${\cal L}_D$
accounts for the dissipation of the cavities (at rate $\kappa_j$) and reads $ {\cal L}_{D}\rho{=}\sum_{j=1}^{2N}\kappa_j(2\hat a_j\rho \hat a_j\da{-}\{\hat a_j\da \hat a_j,\rho\})$. Finally, ${\cal L}_S$ accounts for the driving (at rate $\zeta$)
of the first-end pair of cavities $(1,N + 1)$ by the external two-mode squeezed field~\cite{kraus,paternostro2004,adesso10}:
\begin{equation}
\label{LS}
\begin{aligned}
&{\cal L}_S\rho{=}
2\, \zeta\, \bar m
({\hat a_1\rho \hat a_{N+1}{+}\hat a_{N+1}\rho \hat a_1{-}\hat a_1\hat a_{N+1}\rho{-}\rho \hat a_1\hat a_{N+1}}{+}\text{h.c.})
\\&{+}
\sum_{j=1,N+1}\!\zeta\,[(\bar n{+}1)(2\hat a_j\rho \hat a_j\da{-}\{\hat a_j\da \hat a_j,\rho\}){+}\bar n({2\hat a_j\da\rho \hat a_j{-}\{\hat a_j \hat a_j\da,\rho\}})]\; .
\end{aligned}
\nn
\end{equation}
The sum is over indices $j{=}1$ and $j{=}N+1$ {\it only}, while $\bar n$ and $\bar m$ are related to the statistics of the driving two-mode entangled field: $\bar n$ is the same average photon number for both modes, $\bar m$ accounts for the inter-mode correlations, and
$\bar m{\leq}\sqrt{\bar n(\bar n+1)}$, with equality holding in the squeezed vacuum.
This effective model is based on the elimination of the degrees of freedom of the reservoir (the driving field) in the limit of large squeezing bandwidth~\cite{kraus,paternostro2004,adesso10}. The entanglement in the driving field is the resource to be transferred via the replication mechanism. The state of the driving field is
$\rho\al{in}_{sq}{=}\hat U_{in}\rho_T \hat U_{in}\da$,
with
$\hat U_{in}{=}e^{{\int d\omega r(\omega)\pt{\hat a_\omega\da \hat b_\omega\da-\hat a_\omega \hat b_\omega}}}$
where $\hat a_\omega$ and $\hat b_\omega$ are the field mode operators and $\rho_T$ a thermal state with
$\bar n_T$ average photons. The condition of large squeezing bandwidth corresponds to an
almost constant squeezing parameter, $r(\omega){\sim}r_0$,
over a sufficiently large range of frequencies around the cavity resonance.
In this situation, the parameters characterizing the entangled driving field are
$\bar n{=}\bar n_T{+}(2\bar n_T+1)\sinh^2 r_0,~\bar m{=}(\bar n_T{+}1/2)\sinh (2r_0)$.
The entanglement is quantified by the logarithmic negativity
${\cal E}_N{=}\max[0,-\log\nu_-]$
with $\nu_-{=}2\bar n{+}1{-}2\bar m$ the smallest symplectic eigenvalue of the partially transposed covariance matrix for the two-mode field \cite{review}. The state is entangled iff
$\nu_-{<}1$, which implies $\bar m{>}\bar n$.
\begin{figure*}[t!]
\includegraphics[width=18cm]{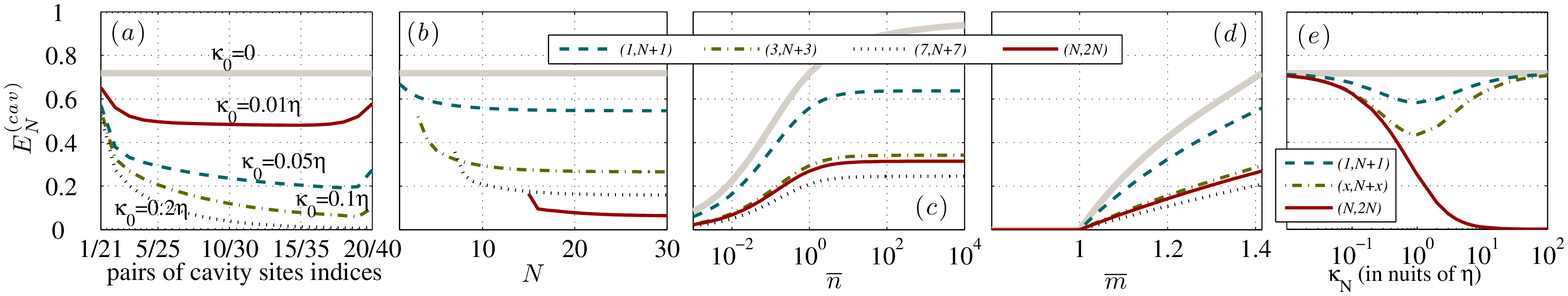}
\caption{
$E_N^{(cav)}$ as a function of (a) the pair-site label $(j,N+j)$ (with $N{=}20$ and $\kappa_j{\equiv}\kappa_0~\forall{j}
$), (b) $N$ (with $\kappa_j{=}0.1\eta,~\forall{j}$), (c) $\bar n$ (with $\bar m{=}\sqrt{\bar n(\bar n+1)}$), (d) $\bar m$ (with $\bar n{=}1$), and (e) $\kappa_N{=}\kappa_{2N}$ (with $\kappa_{j\neq{N,2N}}{=}0$). In all cases $\eta_j=\eta,~\forall j$, and $\zeta{=}\eta$.
The remaining parameters in (a), (b) and (e) are $\bar n{=}1$ and $\bar m{=}\sqrt{\bar n(\bar n+1)}$; in (c), and (d) they are $\kappa_j{=}0.1\eta$ and $N{=}10$.
The insets indicate the pair $(j,N+j)$  corresponding to each line.
In (e) the dash-dotted curve corresponds to all pairs $(j,N{+}j)$ for $j\in\pq{2,N{-}1}$; These results are independent of $N$ and have been verified numerically for arrays of size up to $N=30$.
The solid thick (gray) lines report the entanglement of the driving field which is equal to the entanglement of each pair when $\kappa_j=0,\ \forall j$.
}
\label{fig2b}
\end{figure*}

An exact analytical solution for the steady state is
obtained if the arrays are driven by a two-mode squeezed vacuum ($\bar m=\sqrt{\bar n(\bar n+1)}$), and ${\cal L}_D{=}0$. To obtain the steady state in this situation, we
exploit the squeezing transformation $U{=}\otimes^{N}_{j=1}{ U_{j,N+j}}$, with $ U_{j,N+j}{=}e^{{{(-1)^jr_0 ({\hat a_j\da \hat a_{N+j}\da-\hat a_j \hat a_{N+j}}}})}$, which maps the system into an equivalent one, whose density matrix $\tilde \rho=U\da \rho U$ satisfies the master equation
$\dot{\tilde\rho}{=}{-i} [{ H_c + {\tilde H}_{cs},\tilde \rho}] + {\tilde{\cal L}_S}\tilde \rho\equiv \tilde{\cal L}\tilde \rho$. The new dissipative term reads $\tilde{\cal L}_S \tilde\rho{=}
\sum_{j=1,N+1}\zeta\pq{\pt{2\hat a_j \tilde\rho \hat a_j\da{-}\{\hat a_j\da\hat a_j,\tilde\rho\}}}$, and the transformed Hamiltonian for the cavity-atom interaction is
${\tilde H}_{cs}{=}\sum_{j=1}^{N}g_j[\hat a_j^\dag\hat C_j(\bar n)+\hat a_{N+j}\da\hat D_{j}(\bar n)+{\rm h.c.}]$, with $\hat C_j(\bar n)={\sqrt{\bar n+1}\ \hat \sigma_j+(-1)^j\sqrt{\bar n}\ \hat \sigma_{N+j}\da}$ and $D_{j}(\bar n)={\sqrt{\bar n+1}\ \hat \sigma_{j+N}+(-1)^j\sqrt{\bar n}\ \hat \sigma_{j}\da}$. This shows that, in the new representation, the arrays are in contact with a vacuum reservoir and that each field mode interacts with two atoms at sites $(j,N+j)$. It turns out that, regardless of the actual values of $g_j$ and $\eta_j,\ \forall j\in\pq{1,N}$, the unique steady state is the pure state (which satisfies $\tilde{\cal L}\kb{\varphi}{\varphi}=0$) of the form
$\ke{\varphi}=\ke{\phi}\otimes^{N}_{j{=}1}\ke{\tilde 0,\tilde 0}_{j,N+j}$, i.e. the tensor product of the transformed modes' vacua with the atomic entangled state
\begin{equation}\label{phi}
\begin{aligned}
\ke{\phi}{=}\bigotimes_{j=1}^{N} [{
\sqrt{1-c^2_{\bar n}}\ke{1,1}_{j,N+j}
{+}{(-1)^{j+1}}c_{\bar n}\ke{2,2}_{j,N+j}}] \; .
\end{aligned}
\end{equation}
Here $\ke{1}$ and $\ke{2}$ indicate the ground and excited atomic states, and
$c_{\bar n}=\sqrt{\bar n/(2\bar n+1)}$.
Due to the destructive interference between transitions amplitudes involving the atomic pair $(j,N+j)$ that is coupled to the same mode, state $\ke{\varphi}$ is such that the atoms are decoupled from the field. Moreover, it is not affected by dissipation because the field modes are in their vacuum state. Therefore, during the dynamics, population accumulates, eventually pumping the system into the entangled state Eq.~(\ref{phi}). Going back to the original representation (by inverting the transformation $U$)
also the field modes become entangled in inter-array two-mode squeezed vacua for each pair $(j,N+j)$:
$U_{j,N+j}\ke{\tilde 0,\tilde 0}_{j,N+j}$. All inter-array field-pairs have the same entanglement of the input driving field, thus realizing a {\it perfect entanglement replication mechanism}. On the other hand, the entanglement of all inter-array atomic pairs is the same as that discussed in Refs.~\cite{kraus,paternostro2004,adesso10} for a single atomic pair, but with the essential difference that it is now {\it exactly} replicated across all the $N$ pairs. {\it This is the main result of this Letter}:
from an ideal, infinitely entangled state of the driving field one obtains by engineered dissipation an arbitrary number of EPR field pairs and Bell states of the atomic pairs. In general, the entanglement of the pairs is limited only by the amount of entanglement of the driving field. Moreover, as will be shown below, this result is rather general as it holds valid also for spin chains and arrays of harmonic oscillators.

We will now study the effects of a non-negligible thermal nature of the driving field, and of other sources of dissipation and noise. We consider first the limit in which the model reduces to two chains of harmonic oscillators, i.e. when the atoms are not present ($g_j=0\ \forall j$).
In this case an exact analytical solution is found also if the external field is not perfectly squeezed, $\bar m{\leq}\sqrt{\bar n(\bar n+1)}$. We still assume that ${\cal L}_D = 0$, and we find that, in the squeezed representation, the steady state of each cavity is thermal, $\tilde\rho\al{j}_T$, with mean occupation number $\bar n_T$. In the anti-transformed representation, this corresponds to a two-mode squeezed thermal state for each pair of field modes $(j,N+j)$ that reads:
$U_{j,j+N}\tilde\rho_T\al{j}\otimes\tilde\rho_T\al{j+N} U_{j,j+N}\da$.
The corresponding steady-state entanglement
is {\it the same} as that of the driving field, regardless of $j$, $N$, $\bar n$ and $\bar m$.
Therefore, the exact replication of the driving field entanglement takes place also in this case.
When the other sources of dissipation described by ${\cal L}_D$ are included,
the steady state of the system can be determined numerically, and the logarithmic negativity ${\cal E}_N[j,k]$ of any pair $(j,k)$ of cavity fields is obtained from the corresponding covariance matrix~\cite{review}. Quantitatively, we study the logarithmic negativity normalized to unity,
defined as ${E}^{(cav)}_N[j,k]{=}{ {\cal E}_N[j,k]}/({1+{\cal E}_N[j,k]})$.

\begin{figure}[t!]
\includegraphics[width=8cm]{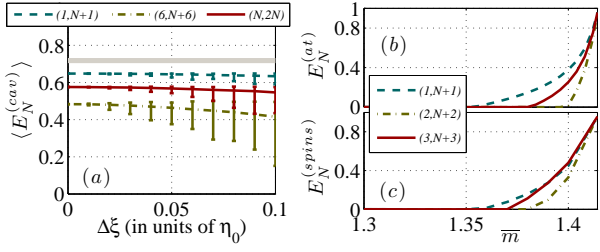}
\caption{
(a) $E_N^{cav}$ for a model with random couplings
as specified in the text.
The curves are obtained averaging the result over $500$ realizations. For each value of $\Delta\xi$ the vertical bars represent the interval between the realizations of maximum and minimum entanglement. The other parameters are $N=10$, $\bar n=1$, $\bar m=\sqrt{\bar n(\bar n+1)}$, $\zeta=\eta_0$, and $\kappa_j=0.02\eta_0 \forall j$.
(b)-(c) Comparison between the logarithmic negativity for atoms in cavity arrays, $E_N^{(at)}$ (b), and for spins in $XX$ spin chains, $E_N^{(spins)}$(c),
as functions of $\bar m$ for $\bar n = 1$.
The remaining parameters are
$\kappa_j{=}0~\forall j$, $N{=}3$,  $\zeta{=}\eta$, and $g{=}0.01\eta$
for the atoms, and $N=3$, and $J_j=\gamma~\forall j$ for the spins.
The insets specify the correspondence between curves and pairs $(j,j+N)$.
}
\label{fig3b}
\end{figure}

Most of the results to follow are obtained for a reservoir with $\bar n=1$, such that the corresponding entanglement is relatively small. Remarkably, even in this strongly non-ideal situation, the replication mechanism is significantly resilient to the added noise.
As shown in Fig.~\ref{fig2b} (a), the entanglement decreases with the decay rate of the cavities.
At fixed decay rate, the largest $E_N^{(cav)}$ is achieved by the pair $(1,N+1)$ that is directly coupled to the driving field. The entanglement of the other pairs decreases moderately with the distance from the driven pair and exhibits a weak revival for a few pairs at the opposite end of the arrays. Fig.~\ref{fig2b} (b) illustrates how the entanglement mildly decays with the size of the arrays, remaining nonvanishing up to large values of $N$. Hence, the entanglement replication mechanism exhibits a notable robustness in the presence of losses. The dependence of the entanglement on the statistics of the input field is shown in Fig.~\ref{fig2b} (c) and (d). When the driving is a squeezed vacuum, its
entanglement increases with $\bar n$ [gray line in Fig.~\ref{fig2b} (c)]
and reaches unity asymptotically as $\bar n{\to}\infty$. For lossy cavities,
the entanglement saturates to a value smaller than unity that depends on the pair
being considered. The entanglement distributed through a squeezed thermal state is reported in
Fig.~\ref{fig2b} (d) showing that $E_N^{(cav)}$ is nonvanishing for all values of $\bar m$
for which the driving field is entangled $(\bar m > \bar n)$. When only the end cavities are open ($\kappa_{j\neq N,2N}=0$), the pairwise entanglement is minimum at $\kappa_N=\kappa_{2N}{\simeq}\eta$ for all pairs $(j,N+j)$ except for pair ($N,2N$) whose entanglement instead decreases
monotonically with $\kappa_N$ [See Fig.~\ref{fig2b} (e)]. As $\kappa_N$ increases the coherent coupling between the last cavity of each array and the neighboring one is progressively inhibited. At large values of $\kappa_N$ each of them is effectively decoupled from the rest of the system, whose entanglement is thus restored to the value of the non-dissipative case. Moreover, the field leaking out of the last pair of cavities is entangled as well \cite{epaps} and even equal to that of the driving field for some frequencies
\cite{epaps}. This feature allows for the re-usability of the transferred entanglement for networking protocols. So far we have discussed results obtained with homogeneous couplings $\eta_j{\equiv}\eta$. Analogous results hold even with intra-array patterns of inhomogeneous couplings, as long as the two arrays remain equal. Asymmetries between the arrays reduce the inter-array entanglement, but the replication mechanism remains valid as long as they are not too strong. This is shown in Fig.~\ref{fig3b} (a), obtained for random couplings $\eta_j=\eta_0+\xi_j,\ {\rm with}\ j\in\pq{1,2N}$, where $\xi_j$ are zero-mean random variables uniformly distributed in a range $\Delta\xi$.

When each cavity interacts with a two-level atom we can study the entanglement properties of the atoms by approximating the system with an effective spin model. We focus on the weak coupling limit, such that the couplings $g_j$ between the atoms and the cavities are sufficiently small~\cite{epaps} and we can adiabatically eliminate the cavity fields to find a closed equation for the atoms. The resulting spin model exhibits non-trivial long-range interactions and collective decay of the spins, as reported in detail in the Supplementary Material~\cite{epaps}. Here we discuss the results relevant for the corresponding steady state. Let us consider the logarithmic negativity $E_N^{(at)}[j,k]=\log_2||{\rho_{jk}^{{\rm PT}}}||_1$
of the state $\rho_{jk}$ of the atomic pair $(j,k)$, where $||\cdot||_1$ is the trace norm and ${\rm PT}$ stands for partial transposition. The entanglement properties of the atoms are similar to those of the free cavity fields. However, at variance with the latter case, $E_N^{(at)}[j,k]$ is sensitive to the statistics of the driving entangled field and decreases more rapidly with decreasing $\bar m$
as illustrated in Fig.~\ref{fig3b} (b).

The effective spin model with long-range interactions can be compared with the case in which two independent spin chains with $XX$ short-range interactions are coupled on one end to the driving field. As shown in Fig.~\ref{fig3b} (b) and (c), one obtains very similar results. The master equation for this case reads
$\dot\rho=-i [{H_s,\rho}]+{\cal L}_S\rho$, with
$ H_{s}{=}\frac12
\sum_{j=1}^{N-1}\sum_{k=0,N}{{J_j}({\hat \sigma_{k+j}^{x} \hat \sigma_{j+k+1}^{x}{+}\hat \sigma_{k+j}^{y}\hat \sigma_{j+k+1}^{y}}})$, where $J_j$ is the spin-spin coupling, and
$\hat\sigma_j^{x,y}$ are the Pauli spin operators. The effect of the driving field is described by
\begin{equation}
\begin{aligned}
&\frac{{\cal L}_S\rho}{\gamma}{=}
2 \bar m({\hat \sigma_1\rho \hat \sigma_{N+1}{+}\hat \sigma_{N+1}\rho \hat \sigma_1{-}\hat \sigma_1\hat \sigma_{N+1}\rho{-}\rho\hat \sigma_1\hat \sigma_{N+1}}{+}\text{h.c})\\
&{+}
\sum_{j=1,N+1}[(\bar n{+}1)({2\hat \sigma_j\rho \hat \sigma_j\da{-}\{\hat \sigma_j\da \hat \sigma_j,\rho\}}){+}\bar n({2\hat \sigma_j\da\rho \hat \sigma_j{-}\{\hat \sigma_j\hat  \sigma_j\da,\rho\}})] \; ,
\end{aligned}
\nn
\end{equation}
with $\sigma_j$ ($\sigma_j\da$) the spin lowering (rising) operator.
While in the cavity-atom system the effective spin-spin interactions are long range~\cite{epaps}, here we deal only with local ones. Nevertheless, entanglement replication continues to hold. Indeed, the stationary state of the system for $\bar m{=}\sqrt{\bar n( \bar n+1)}$ can be evaluated
analytically and coincides with that of Eq.~(\ref{phi}), where $\ke{1}$ and $\ke{2}$
now denote, respectively, the spin up and spin down states. Finally, we observe that the similarity of the steady-state entanglement properties in the two systems holds even when the driving field has a nonvanishing thermal component, as shown in Fig.~\ref{fig3b} (b) and (c). This result shows the generality of the entanglement replication mechanism which is largely independent of the specific physical realization.

In conclusion, we have discussed a scheme realizing the replication of entanglement, based on the interface of a driving two-mode entangled field with two distant and independent dissipative many-body systems. The replication mechanism works efficiently both for arrays of discrete- and continuous-variable systems.
Since the phenomenon occurs in the steady state of the irreversible driven-dissipative dynamics, it exhibits an intrinsic robustness against the detrimental effects of noise.
We have highlighted the roles played by quantum interference and the competition between dissipation, driving, and interactions in producing such a steady state. The corresponding entanglement is robust against deviations from ideal conditions including a nonvanishing thermal component of the driving field, asymmetries between the arrays, and decay of the cavity fields.
Ideally, the replication mechanism yields an arbitrary number of maximally entangled pairs and is scale-free in the sense that it is independent of the actual length of the arrays. Thus, it is a potentially valuable resource for remote quantum communication and distributed quantum computation~\cite{nonlocal,kimble} that could be combined with other driven-dissipative strategies for the realization of scalable quantum networks \cite{vollbrecht}. Seen from a different viewpoint, this scheme implements a protocol of long-distance entanglement distribution~\cite{illubase,illufurther} and nested entangled-pair production~\cite{matryoska},
two key tasks for quantum networking, achieved via the interactions intrinsic in many-body systems.

The outlined scheme is general and flexible enough to find application in many systems which effectively realize chains of harmonic oscillators or spins, such as cavity/circuit-QED~\cite{implementations,circuitQED}, arrays of optomechanical systems, trapped ions, or ultracold atoms in optical lattices. The mechanism could be verified with arrays of coupled resonators, recently produced in photonic crystals \cite{Notomi,Sato}, which realize chains of linearly coupled harmonic oscillators. In Ref.~\cite{Notomi} the cavities are almost resonant and they interact with nearest-neighbor couplings of strength within the range $\sim60-2000$ GHz. These values can be tailored by selecting the distance between the cavities. The reported cavity line-width is of the order of $\sim\!\!1$ GHz. These parameters are consistent with those discussed in our analysis. However, the broadest squeezing at the wavelength of the resonators of Ref.~\cite{Notomi} ($\sim\!\!1.5$ $\mu$m) has a bandwidth of about $\sim\!\!2$ GHz~\cite{Ast}. This value is still relatively small and does not well satisfy the broadband condition assumed throughout our work. Nevertheless, larger squeezing bandwidths and photonic-crystal nano-cavities with weaker decay rates are expected to be realizable in the near future~\cite{Ast,Taguchi}, thus matching the required condition. On the other hand, the currently available experimental situation might already suffice for testing the entanglement replication mechanism. Indeed, a relevant theoretical question, which deserves further investigation, is whether entanglement replication holds also for driving squeezed fields of finite bandwidth.

FI and SZ acknowledge financial support through the FP7 STREP Project HIP, Grant Agreement n. 221889, and iQIT, Grant Agreement n. 270843. GA is supported by a Nottingham Early Career Research and Knowledge Transfer Award. MP acknowledges financial support from the UK EPSRC through a Career Acceleration Fellowship and under the "New Directions for Research Leaders" initiative (EP/G004759/1).

\vspace{0.4cm}

\begin{center} {\bf SUPPLEMENTARY MATERIAL}
\end{center}

\vspace{0.2cm}

\section{Logarithmic negativity of the output field}\label{SCM}

\begin{figure}[b!]
\includegraphics[width=8.5cm]{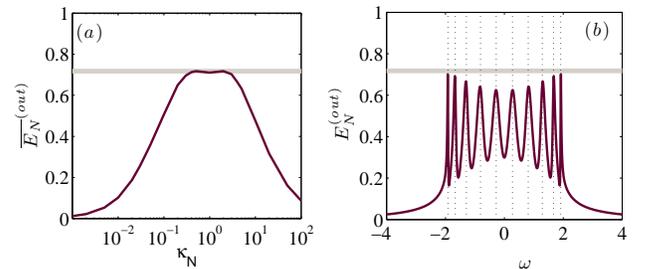}
\caption{(a)
$\overline{E_N}^{(out)}$ as a function of $\kappa_N{=}\kappa_{2N}$ (in units of $\eta$).
(b)
$E_N^{(out)}$ as a function of $\omega$ (in units of $\eta$) for
$\kappa_N=\kappa_{2N}=0.4\eta$.
The frequency $\omega$ is relative to the resonance of the cavities.
The vertical dotted lines indicate the frequency of the normal modes of the arrays.
In both panels the remaining parameters are
$\kappa_{j\neq{N,2N}}{=}0$, $N{=}10$, $\zeta{=}0.5\eta$, $\bar n{=}1$, $\bar m=\sqrt{2}$.
The solid (gray) lines indicate the entanglement of the input squeezed vacuum.}
\label{fig5}
\end{figure}

If the last pair of cavities of the two arrays are open, then
the field leaking out of the last pair of cavities is entangled.
The
 corresponding logarithmic negativity $\overline{E_N}^{(out)}$
[See Fig.~\ref{fig5} (a)]
is maximum at intermediate values of $\kappa_N$. At small values of $\kappa_N$ the number of photons leaking out of the cavities is too low, and so is the associated entanglement. Similarly, at large $\kappa_N$ dissipation is too strong for the build-up of entanglement in the output fields (See below for the detailed evaluation of $\overline{E_N}^{(out)}$). Remarkably, in this situation, $\overline{E_N}^{(out)}$ can reach values very close to those of the driving field,
thus demonstrating the effectiveness of the scheme and the re-usability of the transferred entanglement for networking protocols.

The cavities in the arrays can emit photons into the continuum of modes of the electromagnetic field of the environment. Therefore, the output field is made of a continuum of frequencies.
In order to determine the logarithmic negativity, $E_N^{(\rm out)}(\omega)$, of the
frequency components of the output field,
we have to evaluate the spectrum of the covariance matrix.
Thereafter, the value of $E_N^{(\rm out)}(\omega)$ is obtained by applying
the definition of the logarithmic negativity to the spectral components of the covariance matrix.

An example of $E_N^{(out)}(\omega)$, corresponding to the parameters for which the entanglement of the output field reaches, for some frequencies, a value very close to that of the driving field, is shown in Fig.~ \ref{fig5} (b). Here maxima of the entanglement are found in correspondence of the frequencies of the normal modes of the  arrays. The figure Fig.~ \ref{fig5} (a) illustrates the behavior of  $\overline{E_N}^{out}$ as a function of $\kappa_N$, evaluated in terms of the maximum value of the spectrum $E_N^{(out)}(\omega)$, for each value of the decay rate $\kappa_N$, that is $\overline{E_N}^{out}={\rm max}\pq{E_N^{out}(\omega)}$.

In general, the covariance matrix
of the output field can be expressed as
\begin{eqnarray}\label{Gammaomega}
{\bs \Gamma}\al{out}(\omega){=}\frac{1}{2}\pq{{\bs \Theta}{\bf{\cal A}}\al{out}(\omega){\bs \Theta}^{\rm T}+{\bs \Theta}{\bf{\cal A}}\al{out}(\omega)^{\rm T}{\bs \Theta}^{\rm T} }.
\end{eqnarray}
where the elements of the $4N\times 4N$ matrix ${\bs \Theta}$ are
${\bs \Theta}_{j,k}=\delta_{j,2k-1}+\delta_{j,2k-4N-1}+i\pt{\delta_{j,2k}-\delta_{j,2k-4N}}$,  and ${\bf{\cal A}}\al{out}(\omega)$ is the spectrum of the correlation matrix of the output field operators defined as
\begin{eqnarray}\label{Aomega}
{\bf{\cal A}}\al{out}(\omega)=
\pt{
\begin{array}{cc}
 {\bf A}^{--}_{out}(\omega) & {\bf A}^{-+}_{out}(\omega) \\
 {\bf A}^{+-}_{out}(\omega) & {\bf A}^{++}_{out}(\omega)
\end{array}
}
\end{eqnarray}
with
\begin{eqnarray}\label{A0omega}
\pt{ {\bf A}^{\alpha\beta}_{out}(\omega) }_{j,k}=\int_{-\infty}^\infty \dd t\ \ee^{i\omega t} \av{\hat{a}^\alpha_{j\,out}(t)\ {\hat{a}^\beta_{k\,out}}(0)}_{st},\nn\\
\end{eqnarray}
where $\alpha,\beta\in\pg{+,-}$,
and we use the definitions $\hat {a}_{j\,out}^+\equiv \hat {a}_{j\,out}\da$ and $\hat {a}_{j\,out}^-\equiv \hat {a}_{j\,out}$ for the creation and annihilation operators of the output field~\cite{Gardiner}.

The correlation functions of the output field, in Eq.~(\ref{A0omega}), can be evaluated by means of the input-output formalism~\cite{Gardiner} which allows to express the correlation functions of the output operators in terms of the correlation functions of the system operators.
Exploiting this formalism, one finds
\begin{widetext}
\begin{eqnarray}
{\cal A}\al{out}(\omega)=
-\pt{
\begin{array}{cc}
{\bf K}\pq{({\bf M}^-+\ii\omega\id)^{-1} {\bf A}^{--}_0 +{{\bf A}^{--}_0}^T({\bf M}^--\ii\omega\id)^{-1} }{\bf K}
& {\bf K}\pq{ ({\bf M}^-+\ii\omega\id)^{-1} {{\bf A}^{+-}_0}^T+{{\bf A}^{+-}_0}^T({\bf M}^+-\ii\omega\id)^{-1}}{\bf K}-\id     \\
{\bf K}\pq{({\bf M}^++\ii\omega\id)^{-1} {\bf A}^{+-}_0 +{\bf A}^{+-}_0({\bf M}^--\ii\omega\id)^{-1} }{\bf K}
&{\bf K}\pq{({\bf M}^++\ii\omega\id)^{-1}  {{\bf A}^{++}_0}^T+{\bf A}^{++}_0({\bf M}^+ -\ii\omega\id)^{-1}  }{\bf K}
\nn
\end{array}
}
\end{eqnarray}
\end{widetext}
%
%
where  $\bf K$ is the $2N\times 2N$ diagonal matrix with elements ${\bf K}_{j,j}=\sqrt{\kappa_j}$, the matrices ${\bf M}^\alpha$ are the matrices of the coefficients in the system of equations for the evolution of the averages of the cavity field operators $\frac{\partial}{\partial t}\langle{\hat{a}_j^\alpha}\rangle=\sum_k {\bf M}_{j,k}^\alpha \langle{\hat{a}_k^\alpha}\rangle$, and the elements of the matrices ${\bf A}_0^{\alpha\beta}$ are the steady-state correlation functions of the cavity field operators, defined as
\begin{equation}
{{\bf A}}_{0\,jk}^{\alpha\beta}={\rm Tr}\left[\hat a_j^\alpha \hat a_k^\beta \rho_{st}^{field}\right].
\end{equation}
The elements of the matrices ${\bf M}^\alpha$ are easily evaluated, whereas the matrices ${\bf A}_0^{\alpha\beta}$ can be computed numerically solving the set of equations for the correlation functions  whose form is found using the master equation for the system dynamics in the main text of the
present work.

%


%


\section{Effective spin model for the two-level systems dynamics}\label{Appeffspin}

Here we consider the model described in the main text of the present work, with homogeneous couplings $\eta_j\equiv \eta$ and $g_j\equiv g$ for all $j$.
When the time scale for the two-level system dynamics ($T_{\rm at}$) is much larger then the time scale for the cavity-fields dynamics ($T_{\rm cav}$), then we can adiabatically eliminate the cavity fields, thereby obtaining an effective spin model for the dynamics of the two-level atoms.
The time scale for the atoms dynamics can be estimated as  $T_{\rm at}\sim 1/g\sqrt{\av{n}+1}$, where $\av{n}$ stands for the average cavity-photon number, whereas the time scales for the fields dynamics is determined by the eigenvalues $\{\xi_j\}$  of the matrix ${\bf M}^\alpha$ defined in Sec.~\ref{SCM}. In particular the time scale for the field dynamics is set by the smaller eigenvalue $T_{\rm cav}\sim 1/{\rm min}\pg{\abs{\xi_j}}$.

Henceforth, the master equation for the system dynamics can be rewritten as
\begin{eqnarray}\label{Meq0}
\dot\rho={\cal L}_0\rho+{\cal L}_1\rho,
\end{eqnarray}
where ${\cal L}_0\rho{=}-i[{H_c,\rho}]{+}{\cal L}_S\rho{+}{\cal L}_D\rho$ and ${\cal L}_1\rho{=}-i[{H_{cs},\rho}]$. At lowest order in the atom-field coupling strength, the master equation $\rho_{s}={\rm Tr}_{field}\pq{\rho}$ describing the dynamics of the two-level atoms only takes the form
\begin{equation}\label{effMeq}
\dot\rho_{s}=\int_0^\infty  dt\  {\rm Tr}_{field}\pg{ {\cal L}_1 \ee^{ {\cal L}_0t}  {\cal L}_1 \rho_{st}^{field}\otimes\rho_{s}   },
\end{equation}
where $\rho_{st}^{field}$ is the steady-state of the field in absence of the interaction with the atoms. 
This expression can be recast as
\begin{equation}\label{eff}
\dot\rho_{s}{=}\sum_{j,k=1}^{4N}\pq{ {\bar\sigma}_j\pt{ {\bf{\cal T}}_{k,j}+\bar {\bf{\cal T}}_{j,k} }\rho_s{\bar\sigma}_k
{-}{\bar\sigma}_j{\bf{\cal T}}_{j,k}{\bar\sigma}_k \rho_s{-}\rho_s {\bar\sigma}_j\bar {\bf{\cal T}}_{k,j}{\bar\sigma}_k }
\end{equation}
where ${\bar \sigma}_j\equiv \hat{\sigma}_j\da$ for $j\leq 2N$ and ${\bar \sigma}_j\equiv \hat{\sigma}_{j-2N}$ otherwise.
We have introduced the $4N\times 4N$ matrices ${\cal T}$ and $\bar {\cal T}$ with elements
\begin{equation}\label{TT}
{\bf{\cal O}}_{j,k}{=}-g^2\int_0^\infty \dd t\  {\rm Tr}_{field}\pg{  {\bar a}_j \ee^{{\cal L}_0t}  {\cal R}_k  }~({\cal O}={\cal T},\bar{\cal T})
\end{equation}
with ${\cal R}_k={\bar a}_k \rho_{st}$ (${\cal R}_k=\rho_{st}  {\bar a}_k$) for
${\cal O}={\cal T}$ (${\cal O}=\bar{\cal T}$).
Here
${\bar a}_j\equiv\hat{a}_j$ for $j\leq 2N$ and ${\bar a}_j\equiv\hat{a}_{j-2N}\da$ otherwise.
Both ${\cal T}$ and $\bar{\cal T}$ can be expressed in term of the matrices ${\bf M}^\alpha$ and $ {\bf A}^{\alpha\beta}_0$, introduced above in Section~\ref{SCM} of this supplementary material, as
\begin{equation}
\label{TTT}
\begin{aligned}
{\bf{\cal T}}&=
g^2\pt{
\begin{matrix}
 ({{\bf M}^{-}})^{-1}{\bf A}^{--}_0 & { ({\bf M}^{-}})^{-1}{\bf A}^{-+}_0 \\
 { ({\bf M}^{+}})^{-1}{\bf A}^{+-}_0 & { ({\bf M}^{+}})^{-1}{\bf A}^{++}_0
\end{matrix}
}\\
\bar {\bf{\cal T}}&=
g^2\pt{
\begin{matrix}
 {\bf A}^{--}_0 \ ({\bf M}^{-})^{-1}& {\bf A}^{-+}_0 \  ({\bf M}^{+})^{-1}\\
 {\bf A}^{+-}_0 \  ({\bf M}^{-})^{-1}& {\bf A}^{++}_0 \     ({\bf M}^{+})^{-1}
\end{matrix}
}^{\rm T}.
\end{aligned}
\end{equation}
Equation~(\ref{eff}) and the matrices in Eq.~(\ref{TTT}) have been used for the numerical evaluations presented in the discussion of the atom-cavity model. Eq.~(\ref{eff}) describes a non-trivial spin system where both the spin-spin coherent interactions and the dissipation mechanism can be long-range. An example where such effective spin model can be studied analytically is found for ${\cal L}_D=0$, as seen in the next Subsection.

\subsection{Effective spin model for ${\cal L}_D=0$}\label{effSpins2}

When ${\cal L}_D=0$ the effective master equation takes the form
\begin{equation}\label{effMeq0}
\begin{aligned}
\dot\rho_{s}&=\gamma_\xi\sum_{j,k=1}^{4N}
\pq{ 2\ \bar\sigma_j{\bf{\cal Y}}_{k,j}\al{\xi}\rho_s\bar\sigma_k
-\bar\sigma_j{\bf{\cal Y}}_{j,k}\al{\xi}\bar\sigma_k\rho_s  -\rho_s \bar\sigma_j {\bf{\cal Y}}_{j,k}\al{\xi}\bar\sigma_k  }\\
&-i J\sum_{j,k=1}^{4N}\pq{\bar\sigma_j{\bf{\cal X}}_{j,k}\al{\xi}\bar\sigma_k,\rho_s } \; ,
\end{aligned}
\end{equation}

where $\xi\in\pg{even,odd}$ distinguish between the case in which $N$ is even or odd, the parameters are
\begin{equation}
J=\frac{g^2}{\eta},~~\gamma_{even}=\kappa\frac{ g^2}{\eta^2},~~\gamma_{odd}=\frac{g^2}{\kappa},
\end{equation}
and the $4N\times 4N$ matrices of coefficients ${\bf{\cal X}}\al{\xi}$ and ${\bf{\cal Y}}\al{\xi}$, can be expressed as block matrices
\begin{equation}
\begin{aligned}
&{\bf{\cal X}}\al{\xi}{=}
 \pt{
\begin{array}{cccc}
 && (1+\bar n){\bf X}_\xi &  \\
& & &(1+\bar n){\bf X}_\xi \\
-\bar n{{\bf X}_\xi} & & &\\
&-\bar n{{\bf X}_\xi}& &
\end{array}
}
\\
\label{WW}
&{\bf{\cal Y}}\al{\xi}=\pt{
\begin{array}{cccc}
 & \bar m{\bf W}_\xi& (1+\bar n){\bf Y}_\xi &  \\
\bar m{\bf W}_\xi& & &(1+\bar n){\bf Y}_\xi \\
\bar n{{\bf Y}_\xi} & & &\bar m{\bf W}^*_\xi\\
&\bar n{{\bf Y}_\xi}&\bar m{\bf W}^*_\xi &
\end{array}
}.
\end{aligned}
\end{equation}
Here the missing blocks are null matrices, ${\bf X}_{\xi}$ and ${\bf Y}_{\xi}$ are $N\times N$ matrices whose elements are
\begin{equation}\label{X}
\begin{aligned}
\pt{{\bf X}_{even}}_{j,k}&=\!\!\!\sum_{n,m=1}^{N}\!\!\pt{-1}^{n+1}\pq{\delta_{j,2m}\delta_{j,k+2n-1}{+}\delta_{k,2m}\delta_{j+2n-1,k}},
\nn\\
\pt{{\bf X}_{odd}}_{j,k}&=\!\!\!\sum_{n,m=1}^{N}\!\!\pt{-1}^{n+1}\pq{\delta_{j,2m+1}\delta_{j,k+2n-1}{+}\delta_{k,2m+1}\delta_{j+2n-1,k}},\nn\\
\end{aligned}
\end{equation}
\begin{equation}
\begin{aligned}
\pt{{\bf Y}_{even}}_{j,k}&{=}\sum_{n,m=1}^{N}\pt{-1}^n\pq{\delta_{j,2m}\delta_{j,k+2n}+\delta_{k,2m}\delta_{j+2n,k}}
\nn\\&
+\sum_{m=1}^{N}\delta_{j,2m}\delta_{j,k}\ ,
\nn\\
\pt{{\bf Y}_{odd}}_{j,k}&{=}\sum_{n,m=1}^{N}\pt{-1}^n\pq{\delta_{j,2m-1}\delta_{j,k+2n}+\delta_{k,2m-1}\delta_{j+2n,k}}
\nn\\&
+\sum_{m=1}^{N}\delta_{j,2m-1}\delta_{j,k}\ ,
\end{aligned}
\end{equation}
and  ${\bf W}_\xi =\pt{ {\bf Y}_\xi +i \frac{J}{\gamma_\xi} {\bf X}_\xi }{\bf Z}$ with ${\bf Z}_{j,k}=(-1)^{j-1}\delta_{j,k}$.


The first term in Eq.~(\ref{effMeq0}) describes the coherent interaction between the spins, while
the second one accounts for the dissipation.
The coherent part does not couple spins belonging to different arrays, and the spins in each array are coupled according to the structure defined by the matrix ${\bf X}_\xi$ in Eqs.~({\ref{X}}): the indices of the nonvanishing entries in these matrices correspond to the indices of the coupled spins. The incoherent part, on the other hand, couples both spins from the same array and from different arrays, according  to the pattern defined by the matrix ${\bf{\cal Y}}\al{\xi}$, in Eq.~(\ref{WW}).

\end{document}